\documentclass[twocolumn,twoside,slac_two]{revtex4fj}
\usepackage{graphicx}
\usepackage{fancyhdr}
\bibpunct{[}{]}{;}{a}{}{,}
\def \agile {AGILE}
\def \egret {EGRET}

\def \fermi {{\it Fermi}}
\def \igr {INTEGRAL}
\def \swi {{\it Swift}}

\def \cgro {CGRO}
\def \rxte {RXTE}

\def \ergcmsec{\hbox{erg cm$^{-2}$ s$^{-1}$}}
\def \phcmsec{\hbox{photons cm$^{-2}$ s$^{-1}$}}

\def \gray {$\gamma$-ray}
\def \source {3C~454.3}

\def\aap {Astron. \& Astrophys.\/}

\def\apj {Astrophys. J.\/}
\def\apjl {Astrophys. J. Letters}

\def\mnras {Mon. Not. R. Astron. Soc.\/} 
\def\nat {Nature}

\begin{document}

\title{The gamma-ray flaring properties of the blazar \source{}}

%

\author{S. Vercellone}
\affiliation{INAF--IASF Palermo, Via Ugo La Malfa 153,  90146 Palermo, Italy}
\author{on behalf of the AGILE Team}

%
	\begin{abstract}
%
\source{} is the most variable and intense extragalactic gamma-ray blazar detected by \agile{} and \fermi{} 
during the last 4 years. This remarkable source shows extreme flux variability (about a factor of 20) on a 
time-scale of 24--48 hours, as well as repeated flares on a time-scale of more than a year. The dynamic range, 
from the quiescence up to the most intense gamma-ray super-flare, is of about two orders of magnitude. 
We present the gamma-ray properties of \source{}, comparing both the characteristics 
of flares at different levels and their multi-wavelength behavior. Moreover, an interpretation of both the long- and 
short-term properties of \source{} is reviewed, with particular emphasis on the two gamma-ray super-flares observed 
in 2009 and 2010, when \source{} became the brightest source of the whole gamma-ray sky.
	\end{abstract}

\maketitle

\thispagestyle{fancy}


%
	\section{INTRODUCTION}\label{sec:intro}
%
Multi-wavelength studies of \gray{} active galactic nuclei (AGNs)
date back to the late '70s and early '80s with the COS--B
detection of 3C~273~\cite{Swanenburg78,Bignami81}. 
It was during the '90s, with the launch of \cgro{}, that \egret{}
allowed us to establish blazars as a class of \gray{} emitters and to start
multi wavelength studies of such sources. For a few sources,
it was possible to study both the properties of the spectral energy distributions (SEDs) during
different \gray{} states, and the search for correlated variability
at different bands (e.g., 3C~279~\cite{Hartman2001:3C279:multiwave,Hartman2001:3C279:gammaopt}).
The launches of the \agile{}~\cite{Tavani2009:Missione} and \fermi{}~\cite{Atwood2009:fermi_lat}
\gray{} satellites allowed a tremendous improvement in the monitoring of blazars in the
\gray{} energy band, thanks to their wide field of view and all-sky scanning pointing 
mode\footnote{The \agile{} satellite switched from a pointing to a spinning observing mode on
2009 November because of a permanent reaction wheel failure.}. 

The flat-spectrum radio quasar \source{} (PKS 2251$+$158; $z = 0.859$) became one of the most 
investigated sources, thanks to its high dynamic range flux variations from the radio to the \gray{} 
energy bands. A study~\cite{Vercellone2004:duty} showed that during the \egret{} era this
source exhibited the highest \gray{} activity index 
value\footnote{The activity index $\Psi$ is defined as the ratio
between the number of the source high states and the exposure, in unit of $10^{-7}$\,cm$^{-2}$ s$^{-1}$.
See~\cite{Vercellone2004:duty} for further details.}, 
despite moderate \gray{} peak fluxes ($F_{\rm E>100\,MeV}<2\times10^{-6}$\,\phcmsec{}).
In 2005 May \source{} exhibited an extremely intense flare at different 
energies~\cite{Fuhrmann2006:3C454_REM,Giommi2006:3C454_Swift}, but no information
on the \gray{} flux level was available. 
A simultaneous SED is the ``Rosetta Stone'' which helps to study the emission processes
at different wavelengths.
We obtained almost simultaneous SEDs of \source{} from the radio to the \gray{} energy band
through  dedicated blazar observations both in the low and and in the high emission levels
performed by the WEBT-GASP Consortium~\cite{Villata2004:WEBT:BLLac}  
(radio, optical, and NIR),  and, thanks to its extremely rapid repointing capability,
by \swi{}~\cite{Gehrels2004:Swift}  (optical, UV, X-rays, and hard X-rays).
Here we present the results of four years of \agile{} \gray{} and multi-wavelength observations
of \source{}, discussing its short- and long-term properties.

%
	\section{SHORT-TERM MONITORING RESULTS}\label{sec:short}
%
Since 2007, \agile{} detected several \gray{} flares from \source{}~\cite{Vercellone2008:3C454_ApJ, 
Vercellone2008:3c454:ApJ_P1, Donnarumma2009ApJ_P2, Vercellone2010ApJ_P3, Striani2010ApJ_3c454, 
Pacciani2010ApJ_3c454, Vercellone2011:ApJL_3C454_nov2010}. 
The rapid analysis of the \gray{} data allowed us to provide almost real-time alert to other
Observatories at different wavelengths, from radio up to the hard X-ray. This allowed us to obtain detailed 
multi-wavelength light-curves which can be used to extract time-selected SEDs.
\begin{figure*}[!ht]
\centering
\includegraphics[width=150mm,angle=0]{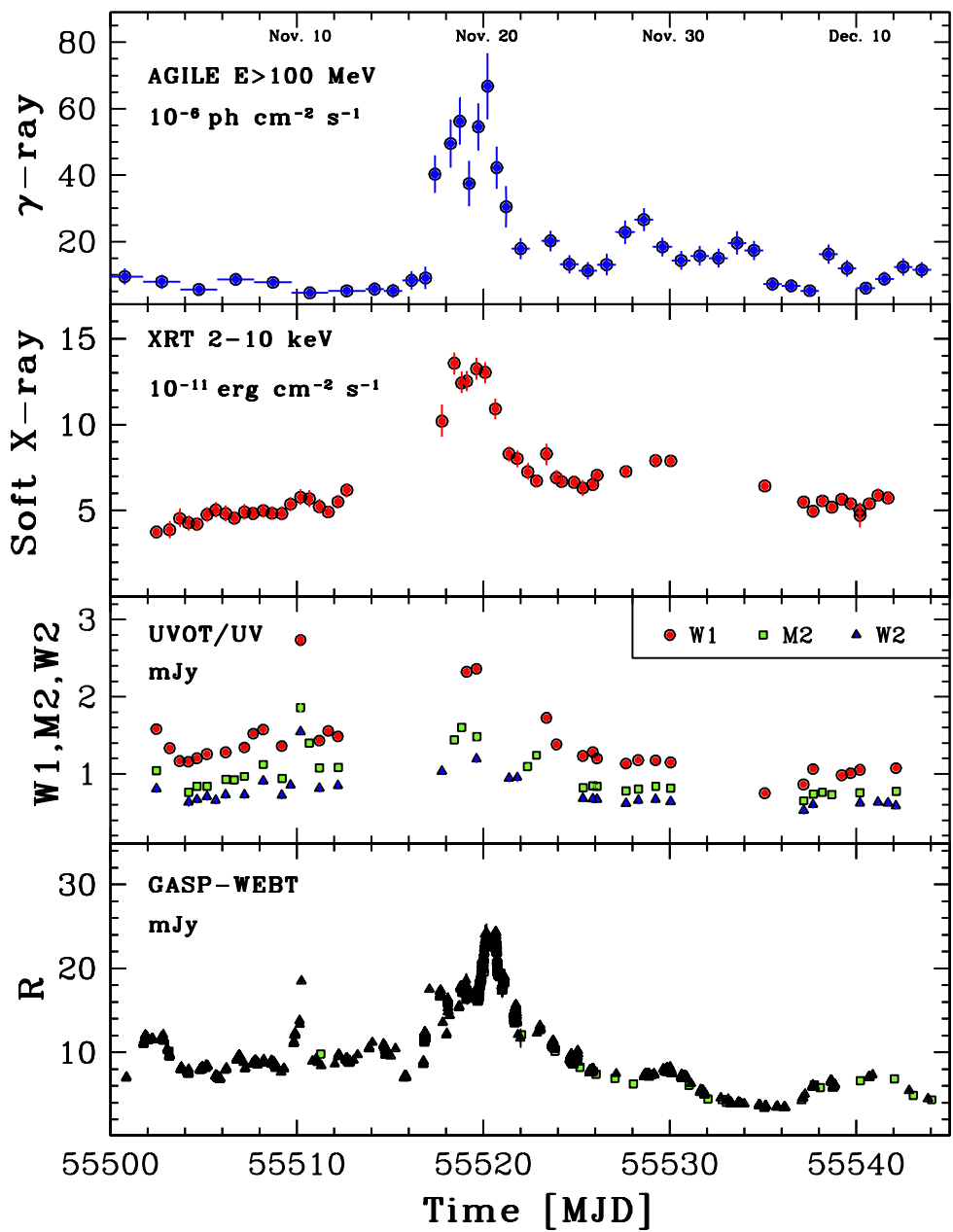}
\caption{
From top to bottom: \agile{} ($E>100$\,MeV), \swi/XRT (2--10\,keV), \swi/UVOT ($w1$, $m2$, $w2$), and
GASP-WEBT ($R$) light-curves obtained during the 2010 November flare. Data 
from~\cite{Vercellone2011:ApJL_3C454_nov2010}.
} \label{fig:f1}
\end{figure*}
Among several \gray{} flares, those which occurred on 2009 December and 2010 November
represent the most intense \gray{} emission from an extra-galactic source so far.
\begin{figure*}[!ht]
\centering
\includegraphics[width=150mm,angle=0]{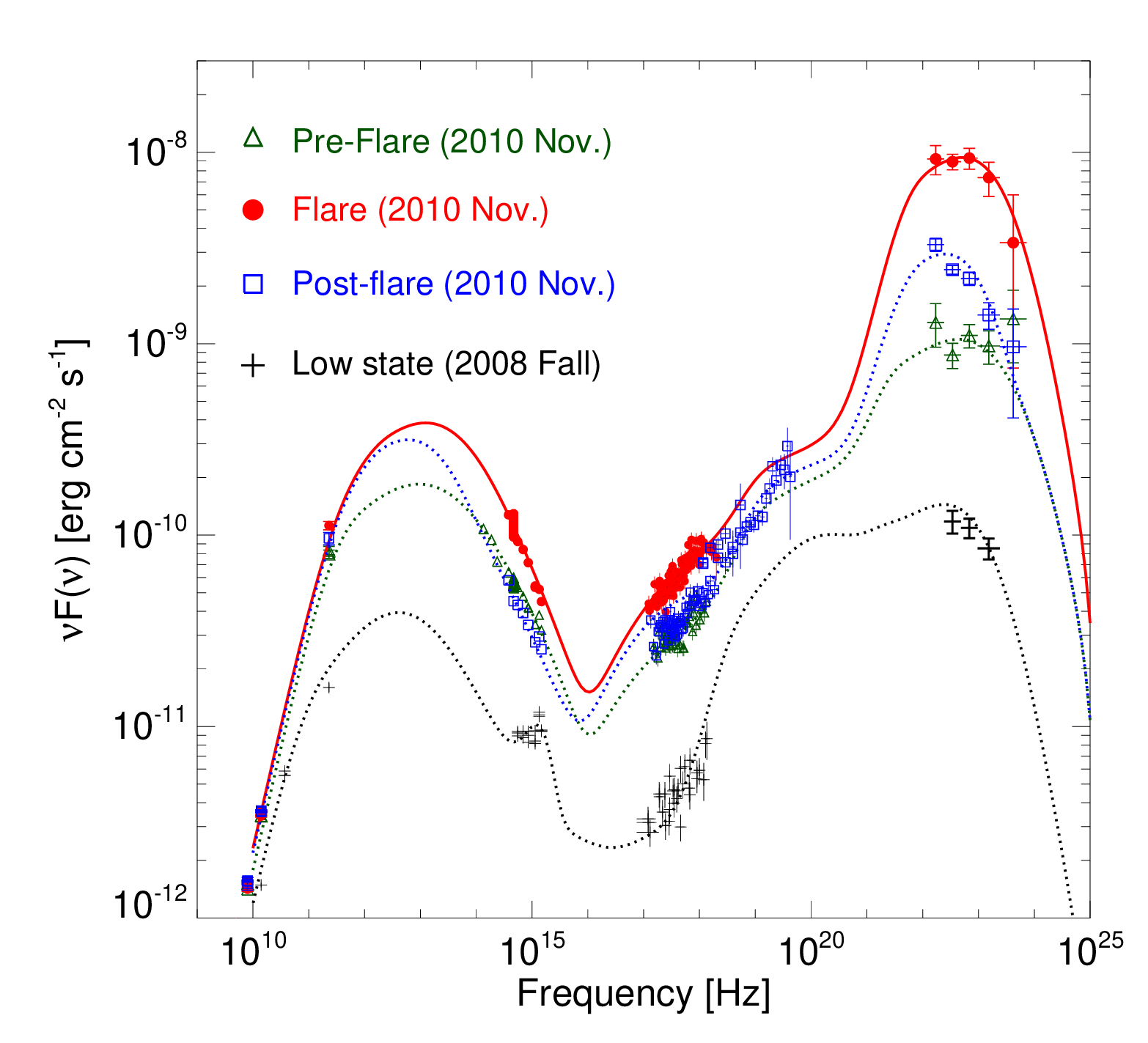}
\caption{
SEDs accumulated during the 2010 November flare (in colors, data 
from~\cite{Vercellone2011:ApJL_3C454_nov2010}) compared with
a SED accumulated during a particularly low \gray{} state in Fall 2008 (in black, data 
from~\cite{Vercellone2010ApJ_P3}).
} \label{fig:f2}
\end{figure*}

The first one occurred on 2009 December 2--3 when \source{}, after an enhanced
\gray{} state lasted about 2 weeks, displayed an extremely fast flux increase up to 
$F_{\rm E>100\,MeV} = (2.0 \pm 0.4) \times 10^{-5}$\,\phcmsec{}, as reported in~\cite{Striani2010ApJ_3c454}.
A multi-wavelength campaign~\cite{Pacciani2010ApJ_3c454} showed that while the
pre- and post-flare SEDs can be modeled in terms of a one-zone synchrotron self-Compton (SSC) plus
external Compton emission (EC), the flare one requires an additional particle component.

The second \gray{} super-flare from \source{} occurred on 2010 November 20, when \source{}
reached a \gray{} peak flux of $F_{\rm E>100\,MeV} = (6.8 \pm 1.0)\times 10^{-5}$\,\phcmsec{}, 
as reported in~\cite{Vercellone2011:ApJL_3C454_nov2010}.
Figure~\ref{fig:f1} shows the multi-wavelength light curves of this extremely intense flare.
While on MJD~55520 a clear peak is visible in all the energy bands, about ten days earlier a remarkably
fast optical flare (with rise and fall of about a factor of 2--2.5 in about 48 hr) occurred, with no
significant variation in the \gray{} energy band and a modest 20\% increase in the X-ray.
Figure~\ref{fig:f2}  shows the SEDs
accumulated before, during, and after the 2010 November \gray{} flare (color points) in comparison with
the SED accumulated in the Fall 2008 during a particularly low state. A modeling of the evolution of the
super-flare SED, taking into account the ``\gray{} orphan'' optical flare on MJD~ 55510, challenges
a model with a uniform external photon field. Moreover, the modeling places the \gray{} emission region
within the BLR.

Short-term observing campaigns allowed us to study the variability properties of \gray{} flares.
\source{} shows extremely fast \gray{} flares (e.g.,~\cite{Foschini2011:fsrq_variab}), which
favor the hypothesis of a dissipation region within the BLR~\cite{Bonnoli2011:3c454}.
Moreover, during previous multi-wavelength campaigns, by combining the \agile{} ($E>100$\,MeV) 
and WEBT-GASP ($R$-band) data we investigated the possible delay between the \gray{} flux variations
with respect to the optical ones. We found~\cite{Vercellone2008:3c454:ApJ_P1, Donnarumma2009ApJ_P2,
Vercellone2010ApJ_P3} a moderate correlation with no time-lag between the emission in the
two energy bands.

%
	\section{LONG-TERM MONITORING RESULTS}\label{sec:long}
\begin{figure}[!ht]
\centering
\includegraphics[width=65mm,angle=-90]{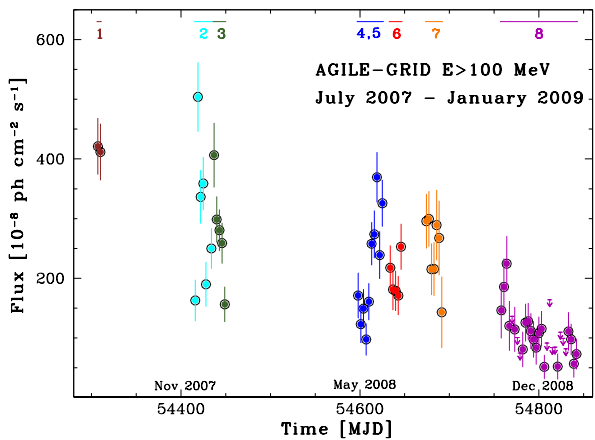}
\caption{
Light-curve for $E>100$\,MeV (time-bin 3d) accumulated during the period 2007 July - January 2009.
Colors and numbers refer to different observing campaigns carried out when \agile{} was observing
in pointed mode. Data from~\cite{Vercellone2010ApJ_P3}.
} \label{fig:f3}
\end{figure}

\begin{figure}[!ht]
\centering
\includegraphics[width=80mm,angle=0]{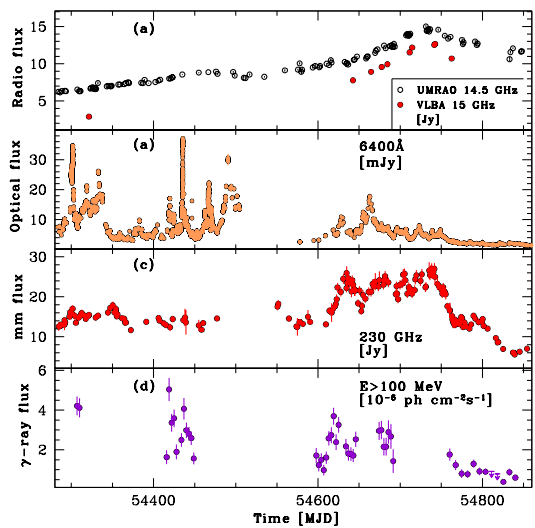}
\caption{
Light-curves at 14.5 and 15.0~GHz (panel a), 6400\AA (panel b), 230~GHz (panel c) and $E>100$\,MeV
(panel d). Data from~\cite{Vercellone2010ApJ_P3}. The time range is the same as in Figure~\ref{fig:f3}.
} \label{fig:f4}
\end{figure}

%
The \agile{} pointing scheme changed since 2009 November, from a pointed strategy to a
quasi all-sky one. Figure~\ref{fig:f3} shows the light-curve accumulated during an eighteen-month
period (2007 July - 2009 January) obtained by combining several pointed observations. We can
appreciate the dynamic range of the different flares (about a factor of 2--4) and the dimming trend 
towards the end of 2008.
Such a long-time coverage was supported by co-ordinated multi-wavelength observations. In particular,
we obtained data with \swi{}, \igr{}, \rxte{}, and from ground-based Facilities (e.g., GASP-WEBT,
UMRAO, the MOJAVE Project).
Figure~\ref{fig:f4} shows, from top to bottom, the light-curves in the radio, optical, millimeter and \gray{} 
energy bands, respectively.

When comparing the radio flux variation with respect to the optical and \gray{} ones, we see how the
former one has a single flux density peak on MJD~54742. This variability is not well correlated with the 
variability at higher frequencies: optical and \gray{} data show more different flares
in the period MJD~54300--54800. Moreover, the radio flux density increase is smooth and longer in time, 
while \gray{} and optical flares are evolving faster. We can speculate that a multiple
source activity in the optical and \gray{} bands is integrated in the radio emitting region in a single event.

The comparison of the light-curves in the mm, optical and \gray{} energy bands can shed some light
on the geometrical properties of the jet. Following~\cite{Villata2009:3C454:GASP:accep}, a possible 
interpretation~\cite{Vercellone2010ApJ_P3} arises in the framework of a change in orientation 
of a curved jet, yielding different alignment configurations  within the jet itself.
During 2007, the more pronounced fluxes and variability of the optical
and \gray{} bands seem to favor the inner portion of the jet as the
more beamed one. On the other hand, the dimming trend in the optical
and in the \gray{} bands, the higher mm flux emission and its enhanced
variability during 2008, seem to indicate that the more extended
region of the jet became more aligned with respect to the observer
line of sight.

\begin{figure}[!ht]
\centering
\includegraphics[width=80mm,angle=0]{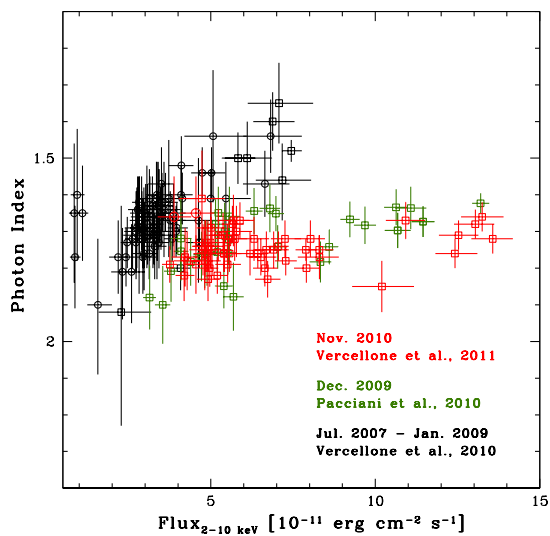}
\caption{
The X-ray photon index as a function of the 2--10\,keV flux during different observing campaigns. 
Data from~\cite{Vercellone2010ApJ_P3,Pacciani2010ApJ_3c454, 
Vercellone2011:ApJL_3C454_nov2010}.
} \label{fig:f5}
\end{figure}

The \swi{} satellite flexibility allowed for the first time to monitor high-energy blazars
during \gray{} flares. Figure~\ref{fig:f5} shows the X-ray photon index as a function of
the 2--10\,keV flux during different observing campaigns.
During the 18-months \agile{} campaign (black symbols), \cite{Vercellone2010ApJ_P3} found 
a clear harder-when-brighter trend, in particular for fluxes above (1--2)$\times 10^{-11}$\,\ergcmsec\,. 
As reported in~\cite{Vercellone2011:ApJL_3C454_nov2010}, during the 2010 November flare (red
points) this trend no longer stands.
Remarkably, \cite{Donnarumma2011:intergral_PoS} show that a behavior similar to the November 2010 one
was already present during the 2009 December \gray{} flare (green points; 
see also~\cite{Pacciani2010ApJ_3c454}).
We can describe the harder-when-brighter trend in terms of a dominant contribution
of the EC off the disk seed photons, EC(Disk), over the SSC component, probably
due to an increase of the accretion rate. We note that an increase
of the electron density ($n_{\rm e}$) and/or of the break energy Lorentz factor ($\gamma_{\rm b}$) 
would cause a softer-when-brighter trend, inconsistent with our findings.
\begin{figure*}[!ht]
\centering
\includegraphics[width=135mm,angle=-90]{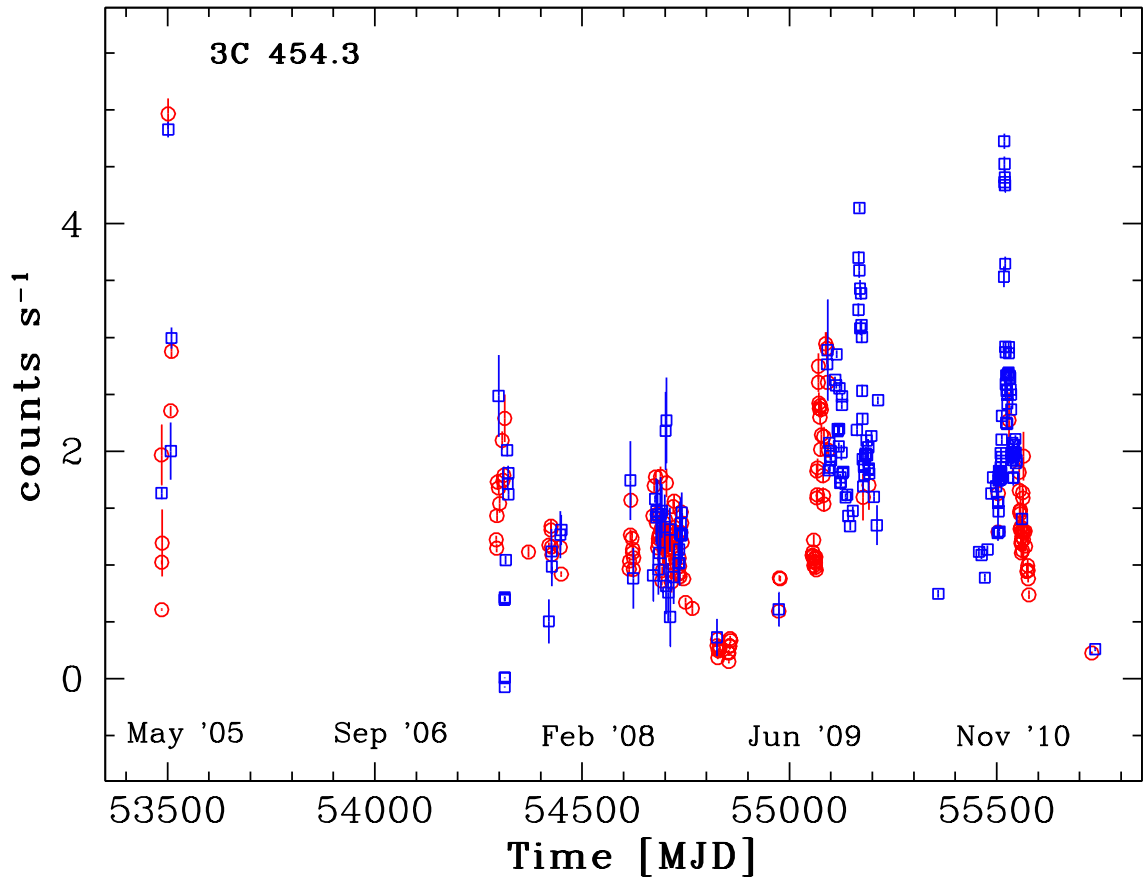}
\caption{
\source{} \swi{}/XRT automatic analysis light-curve (see~\cite{Evans2009:lightcurves}) accumulated during 
the period 2005--2011. Red circles represents Photon Counting mode data while blue squares represent
Windowed Time mode data.
} \label{fig:f6}
\end{figure*}
The constant X-ray photon index during the extreme \gray{} flares in 2009 and 2010 can be 
interpreted in terms of a balance of the SSC contribution with respect to the EC(Disk).
If we assume that $\gamma_{\rm b}$ increases significantly with respect to the 2007--2008
ones ($\gamma_{\rm b} = 200-300$ in 2007--2008, see~\cite{Vercellone2010ApJ_P3}; 
$\gamma_{\rm b} = 700-800$  in 2009--2010, see~\cite{Pacciani2010ApJ_3c454, 
Vercellone2011:ApJL_3C454_nov2010}),
we obtain both an increase of the EC(Disk) component (and the shift of the peak of its
emission to higher frequencies), and a simultaneous increase of the SSC. The net result
is a roughly achromatic increase of the X-ray emission.

%
	\section{DISCUSSION AND CONCLUSIONS}\label{sec:disc_concl}
%
\agile{} and \fermi{} are providing an enormous step forward the understanding of the \gray{} blazar
class, under complementary points of view (the former more prone to dedicated multi-wavelength campaigns
on a small number of sources, the latter discovering and analyzing more than one order of magnitude more
blazars with respect to EGRET). The \gray{} energy band alone cannot conclusively derive the
physical properties of these sources. While radio and optical monitoring has been performed for several
years, only the \swi{} satellite can now provide a detailed monitoring from the optical to the hard X-ray 
energy bands. 

Figure~\ref{fig:f6} shows the \swi{}/XRT light-curve accumulated during the period 2005--2011. Such
a coverage is astonishing because it is obtained with a ``pointed'' instrument, not with an all-sky one.
We can clearly identify the 2005 May, 2009 December, and 2010 November flares. Starting from
2007 mid July (MJD $\approx 54300$) we  see a much denser monitoring of \source{} by means of
\swi{}/XRT, being the epoch of the first \gray{} flare detected by \agile{} (see~\cite{Vercellone2008:3C454_ApJ}).
Moreover, for bright sources ($F_{\rm 2-10\,keV} \ge 5 \times 10^{-11}$\,\ergcmsec,
i.e.,  $\ge$\,1--2\,counts\,s$^{-1}$) we can 
obtain spectral informations on a time-scale of about 5\,ksec, allowing the study of the 
synchrotron-to-inverse Compton transition region. The optical-UV filters on-board \swi{}/UVOT
allow us to study the disc emission in flat-spectrum radio quasars. The black SED in Figure~\ref{fig:f2}
clearly shows the thermal contribution of the accretion disc at $\nu \approx 10^{15}$\,Hz during a particularly
low \gray{} state. This feature is completely swamped out by the intensity of the synchrotron continuum
during flaring \gray{} states (SEDs in colors).

The combination of the \agile{}, \fermi{}, \swi{}, and radio data will allow us to investigate
the properties of the jet. Moreover, simultaneous optical/UV
spectroscopy of the BLRs and optical polarimetry observations would greatly improve our knowledge
of the flaring episodes in \source{}.

\bigskip 
	\begin{acknowledgments}
We acknowledge financial contribution from the agreement ASI-INAF I/009/10/0.
The AGILE Mission is funded by the Italian Space Agency (ASI) with
scientific and programmatic participation by the Italian Institute
of Astrophysics (INAF) and the Italian Institute of Nuclear
Physics (INFN). We acknowledge ASI contract I/089/06/2.
This work made use of data supplied by the UK \swi{} Science Data Centre at the University of Leicester.
\end{acknowledgments}

\bigskip 


\end{document}